\documentclass[11pt,a4paper]{article}
\usepackage{jheppub}
\usepackage{amsmath,amstext,amssymb}
\usepackage{graphicx}
\usepackage{feynmp}
\usepackage{epstopdf}
\usepackage{caption}
\usepackage{float}
\usepackage{booktabs}
\pdfoutput=1

\def\be{\begin{equation}}
\def\ee{\end{equation}}
\def\ba{\begin{eqnarray}}
\def\ea{\end{eqnarray}}
\def\nn{\nonumber}

\usepackage{graphics}
\usepackage{graphicx}
\usepackage{dcolumn}
\usepackage{bm}
\usepackage{epsfig}
\usepackage{graphicx}
\usepackage{multirow}
\usepackage{dcolumn}
\usepackage{graphicx,epsfig}

\begin{document}


\title{\center Emergent symmetries at criticality in multi field RFT/DP }

\author[a]{Jochen Bartels,}
\author[b]{Carlos Contreras,}
\author[c]{Gian Paolo Vacca}

\affiliation[a]{II. Institut f\"{u}r Theoretische Physik, Universit\"{a}t Hamburg, Luruper Chaussee  149,\\
D-22761 Hamburg, Germany}
\affiliation[b]{Departamento de Fisica, Universidad Tecnica Federico Santa Maria, Avda. Espana 1680, Casilla 110-V, Valparaiso, Chile }
\affiliation[c]{INFN Sezione di Bologna and  DIFA, via Irnerio 46, I-40126 Bologna, Italy}

\emailAdd{jochen.bartels@desy.de}
\emailAdd{carlos.contreras@usm.cl}
\emailAdd{vacca@bo.infn.it}

\abstract{ We study multifield extensions of Reggeon Field Theory (also equivalent to Directed Percolation model) at criticality in the perturbative $\epsilon$-expansion below the upper critical dimension $D_c=4$ at one loop, for the special case when all fields have the same scale (anomalous) dimensions. Analyzing all the fixed points of the renormalization group flow for $N=2$ flavors and some for $N=3$, we find scale invariant solutions which are characterized by specific emergent global symmetries of the interacting potential. 
We also study two infinite families as a function of $N$ having $\mathbb{Z}_N$ and $S_{N+1}$ symmetries.
}

\maketitle

\section{Introduction}

In the formulation and study of phenomena in critical and almost-critical physical systems quantum and statistical field theories serve as an important and effective framework. Several aspects of both high energy physics of particles with fundamental interactions and low energy physics of systems subject to quantum/statistical fluctuations can be succesfully described, and even useful descriptions beyond physics towards interdisciplinary domains can be constructed. The most useful paradigm in this framework is given by the renormalization group which, focusing on the description of a system at different scales, permits to link different fundamental properties of a system and in particular scaling, universality and symmetry. Typical systems which are investigated undergo a continuous second order phase transition at criticality, where the correlation lengths become infinite. In many critical systems studied scale invariance is believed to be lifted to conformal invariance (proof of that being available only in rare cases) making available other means of investigations
such as the conformal bootstrap in case of unitary theories. 
The most famous and successful example of this is the QFT description of the Ising model in three dimensions.
On the other hand there are systems of interest which are not conformal at criticality, in particular the ones presenting anisotropic scaling, which are simply scale invariant.

In this work we focus on quantum/statistical models of this kind and in particular on extensions of the so called Reggeon Field Theory,
which was formulated more that sixty years ago during the quest of understanding high energy scattering of strongly interacting particles. After the formulation of the theory~\cite{Gribov:1967vfb,Gribov:1968uy} several works were already devoted to the perturbative investigation of its critical properties~\cite{Abarbanel:1973pq,Sugar:1974td}. In this theory  the role of time is played by the rapidity while the space is the transverse space with respect to the scattering direction.
Even if nowadays strong interactions are successfully described by QCD as a fundamental asymptotically free theory,
still one expects that high energy scattering in the Regge limit could be effectively described by some kind of RFT derived from QCD.
This idea is difficult to be investigated because of the need of a non perturbative analysis, which has led to first attempts based on the use of an exact functional renormalization group (FRG) framework~\cite{Bartels:2015gou,Bartels:2018pin}. 
This QFT is interesting also because it was shown~\cite{Cardy:1980rr} to describe completely different physical phenomena in the framework of non equilibrium phase transitions in statistical physics, known as Directed Percolation (DP). For a general introduction to the subject and this paradigmatic model see~\cite{HHL}. Such equivalence with models of non equilibrium physics has increased the interest in this theory~\cite{Canet:2003yu}, with also a recent perturbative analyses up to three loops~\cite{Adzhemyan:2023wbz}. All the perturbative analyses are typically performed using dimensional regularization in the $\epsilon$-expansion below the upper critical dimension.

We remind here the form of a single field RFT, which is described by the action
\be
S=\int d\tau d^D x \Big[
\psi^{\dag} \left( i\partial{\tau} + \alpha' \nabla^2 + \alpha' \mu \right) \psi
 -\frac{i}{2} \alpha'  \lambda \psi^{\dag}\psi \left(\psi+\psi^\dag\right)  \Big]\nonumber \,,
\ee
where the fields live in $1+D$ dimensions and imaginary time is related to the rapidity. Here $\alpha'$ is the slope and 
$\mu$ is the intercept minus one of the Regge trajectory.
In high energy scattering the physical (transverse) space dimension $D$ is two, 
while for Directed Percolation also other dimensions are of interest.
The dynamics respects the so called projectile-target symmetry in the scattering, or in other words the rapidity-reversal symmetry.

In high energy scattering quantum properties of the particles involved imply the need to have several reggeon fields in the description.
First even in the QCD Pomeron analysis one may encounter a set of Pomeron states which will eventually translate in an effective multifield RFT. More in general the dominant contributions at high energies are given by the so called Pomeron and the Odderon. These objects have their perturbative counterpart in simplified analysis in QCD~\cite{Fadin:1975cb,Balitsky:1978ic} and ~\cite{Bartels:1999yt}, respectively, but in full interacting QCD theory is difficult to formulate a reliable description.
Because of that an idea of considering an effective RFT for these two fields with all the possible specific interactions was proposed and the behavior at criticality  studied both pertubatively~\cite{Bartels:2016ecw, Braun:2023vos} and non perturbatively, with FRG methods~\cite{Bartels:2016ecw}. It is not yet known if this model can have a physical interesting realization in non equilibium statistical physics generalizing the Directed Percolation model. 
Neverthelss one can find in the literature analyses~\cite{tauber1,Janssen:2000ds} related to multispecies populations close to criticality, generalizing the DP universality class. This goes in the direction that we want to investigate in this work.

We shall consider here a multi-field extension of Reggeon Field Theory (RFT) and Directed Percolation models with $N$ fields flavors (reggeons or order parameters/species), focusing on specific families of theories which, allowing real rotations in the flavor space, maintain a global $O(N)$ symmetry at criticality in the two point function, i.e. the non interacting part of the action. 
Such a choice corresponds in RFT to the picture of having $N$ interacting reggeons with the same Regge trajectories.
We shall allow a completely general  interacting critical potential and see how the O(N) symmetry is broken at criticality to some smaller, typically discrete, symmetry so that there will exist a orbit in the space of potentials (couplings) invariant under a subgroup of $O(N)$ of physically equivalent scale invariant theories. This kind of analysis has  been performed for example in an exaustive way, eventually with the help of numerical techniques, in the search of fixed point in interacting standard multi-field scalar theories up to $N=3$ fields ~\cite{Osborn:2017ucf,Codello:2019isr} for cubic interactions and up to $N=4$ fields~\cite{Osborn:2017ucf, Codello:2020lta} with quartic interactions, with further searches pushed up to $N=7$ fields~\cite{Osborn:2020cnf}.

We shall restrict our analysis to the one-loop perturbative $\epsilon$-expansion, which is nevertheless enough to grasp the main qualitative features of these systems.

Let us stress that the set of theories studied here are covering only a part of the possible multi-field theories: 
for example the Pomeron-Odderon system does not satisfy the constraints adopted here, since the two reggeons at criticality have been shown 
to have different slopes, both in the one loop approximation and in non perturbative settings as well. Indeed the system that we investigate in this work must satisfy additional constraints on top of the fixed point conditions of the RG flow.

The symmetries of the scale invariant theories found, can be considered emergent in the sense that we have not imposed them from the start but they were found as the ones characterizing the quantum scale invariant solutions.

Outline of the work:
In Section~\ref{sec2} we start by defining the set of multi-field theories we want to investigate, discussing some of the features
which we need to take into account during the study.
Section~\ref{sec3} is the core of this work and contains the results of our investigations. After discussing some general strategy we present an analysis of the scale invariant theories for $N=2$, adding some details in the discussion of the polynomial invariants, build from the couplings of the potential, under $O(2)$ transformations. This is followed by a partial analysis of the case with $N=3$ fields, which is much more involved. One of the important features we analyze are the global symmetries of the scaling invariant solutions. This allows us from what found for $N=2$ and $N=3$ to guess the form of two families of solution for a generic number $N$ of fields. These two families are then generally investigated in Sections~\ref{SNplus1sol} and~\ref{ZNsol}.
After our conclusions two Appendices are given. In the first Appendix ~\ref{oneloopcomp} we collect the derivation of the perturbative RG flow equations and conditions used in the analysis of Section~\ref{sec3}. In the second Appendix~\ref{qtensors} one can find some definitions and relations used to study in Section~\ref{SNplus1sol} one of the infinite families for general $N$.

\section{Multi-field RFT/DP model}\label{sec2}
Let us define here the theory which we want to investigate. 
We shall consider a set of fields and their conjugate $(\psi_i,\psi_i^\dag)$, for $i=1, \cdots, N$,
whose evolution and interactions are defined by an action of the form
\ba
&{}&S=\int d\tau d^D x \Big[ \sum_j
(\psi_{j}^{\dag}) \left( i\partial{\tau} + \alpha' \nabla^2 + \alpha' \mu \right) \psi_{j} -V(\psi,\psi^\dag) \Big] ,
\ea
where
\ba
V(\psi,\psi^\dag)=
\sum_{jkl} \frac{i}{2} \alpha'  \lambda_{j,kl} \left( \psi_{j}^{\dag}\psi_{k} \psi_{l}+\psi_{j} \psi_{k}^{\dag} \psi_{l}^{\dag} \right) \,.
 \label{action}
\ea
Here $\psi_i$ is an element of the vector $\psi$. In general $\alpha'$ (the so called regge slope) and  $\mu$  (mass parameter related to the intercept) can be real symmetric matrices (with one of them always diagonalizable), but in this analysis we shall consider the case for which all the fields have the same properties, i.e. the kinetic and mass terms of the action enjoy a full $O(N)$ symmetry in flavor space (requiring this property to be true even with the interactions turned on) and $\psi$ and $\psi^\dag$ transform as vectors under rotations.  
Other cases for which such a symmetry is not present from the start are of course physically relevant and  an example in RFT 
is given by an already mentioned system of two reggeons, i.e. a Pomeron interacting with an Odderon~\cite{Bartels:2016ecw, Braun:2023vos}.

In the present analysis we therefore make also a further choice: we enforce the conditions ${\cal C}$ that radiative quantum/statistical corrections do not spoil the $O(N)$ invariance of the quadratic part of the effective action at criticality. 
In this way the full $O(N)$ transformation on the field vector, $\psi' = R_r \psi$ and $\psi^\dagger{}' = R_r \psi^\dagger$ 
for $r \in O(N)$, induces in general a corresponding transformation on the interacting potential $V=\lambda^T P(\psi,\psi^\dagger)$, where $\lambda$ is a vector of couplings and $P$ is a vector of all possible monomial composite operators of the fields. 
Indeed the action of $r \in O(N)$ induces a transformation $P' = \tilde{R}_r P$ and therefore $\lambda' = \tilde{R}^T_r \lambda$ .
One can also decompose this space  of couplings $\lambda$ according to the irreducible representations of $O(N)$ and also construct invariants of different order in the couplings. We shall briefly discuss this for the specific $N=2$ case in Section~\ref{Invariants}.

The goal of this work is to investigate with renormalization group techniques the possible critical theories, 
obtained from the fixed points as zeros of the beta functions (and  satisfy also the additional constraints ${\cal C}$)
which are collected in a vector transforming also  in the same irreducible representation of $O(N)$ as the coupling vector.

Each scale invariant solution, with couplings $\lambda_*$, is characterized by a global symmetry defined in terms of a subgroup $G\subset O(N)$ which leaves the interacting potential invariant, i.e. $\lambda_* = \tilde{R}_g \lambda_*$ for $g \in G$. It is then natural, considering the action of an $O(N)$ transformation, to define a coset $O(N)/G$ which characterizes an orbit in the space of the couplings $\lambda$ (potential $V$) with the same physical properties 
(same universal critical exponents).

Therefore the framework we consider is also interesting from the theoretically point of view since we
show that any scale invariant solutions may have an {\it emergent} symmetry, typically realized by some finite point group $G\subset O(N)$.

The renormalization group analysis is based on the study of the beta functions $\beta_\lambda$ which are derived in Appendix~\ref{oneloopcomp}, together with the anomalous dimensions and the constraints ${\cal C}$ needed to preserve the $O(N)$ symmetry in the part of the action quadratic in the fields. In particular we have:
 the anomalous dimensions 
\ba
\gamma_{ij}&=&  -\frac{1}{8}  \frac{1}{(4\pi)^2}\lambda_{i,lm} \lambda_{j,lm} = \gamma \, \delta_{ij}
\ea
and
\ba
\tilde{\zeta}  &=& \frac{1}{N}  \frac{1}{8(4\pi)^2}\lambda_{i,lm} \lambda_{i,lm} = -\gamma
\ea
together with $\kappa_\perp=-3\gamma$ and the beta functions
\ba
\!\!\!\!\!\!\!\!\!\!\! \beta_{\lambda_{i,jk}}=\beta_{i,jk} &=& -\frac{\epsilon}{2} \lambda_{i,jk} + \tilde{\zeta} \lambda_{i,jk} + 3 \gamma \lambda_{i,jk}
+\frac{1}{2}  \frac{1}{(4\pi)^2}  \left(    \lambda_{i,ab}  \lambda_{j,ac} \lambda_{b,ck} + \lambda_{i,ab}  \lambda_{k,ac} \lambda_{b,cj} \right)\,.
\ea
These equations have to be completed adding the relations ${\cal C}$ of Eq.~\eqref{conditions} which imply the special form of the anomalous dimension matrix $\gamma_{ij}$.

We also introduce the stability matrix $M_{pq}= \frac {\partial \beta_{\lambda  p}}{\partial \lambda_q}$  whose eigenvalues give the universal and invariant critical exponents which characterizes the stability properties of a fixed point, following the evolution under the RG flow after having added a perturbation.

Les us count in general the number of all possible couplings associated to cubic interactions and the number of relations that we have.
For general $N$ we have $\frac{N^2(N+1)}{2}$ different couplings and the same number of beta functions, which are cubic polynomials in the couplings. Moreover we have $(N-1)+\frac{N(N-1)}{2}=\frac{(N+2)(N-1)}{2}$ quadratic contraints ${\cal C}$ coming from the requirement of the existence of a single quadratic invariant making all the field anomalous dimension equal, which are given in Eq.~\eqref{conditions}. 
Finally the continuous $SO(N) \subset O(N)$ symmetry has  $\frac{N(N-1)}{2}$ generators, which can imply that the $O(N)$ symmetry reduces the number of independent physically couplings, since the orbits generated are corresponding to the same physics. 
Let us summarize these numbers for $N$ up to $4$:
\ba
\left( \begin{array}{cccc} 
N \,\,&  \# {\rm couplings} & \,\, \# O(N) {\rm generators} \,\,& \# {\rm constraints} \\
 1 &1 &0&0\\
2 & 6 &1&2\\
3 &18&3&5\\
4 &40&6&9 
\end{array} \right)
\ea

We finally remind the one loop results for a single reggeon (Pomeron), i.e. $N=1$, 
which has a single coupling and just one scaling solution,  fixed point of the beta function:
\ba
\beta_\lambda= -\frac{ \epsilon}{2} \lambda + \frac{3}{4} \frac{\lambda^3}{(4\pi)^2}.
\ea
At the  fixed point one has the following values for the coupling, anomalous dimensions and critical exponents:
\ba
\frac{\lambda_*^2}{(4\pi)^2} = \frac{2}{3} \epsilon \, ,\quad \gamma=-\frac{1}{12} \epsilon \, , \quad \tilde{\zeta}= \frac{1}{12} \epsilon \,, \quad
\nu_\perp=\frac{1}{2}+\frac{1}{16} \epsilon \, , \quad
\frac{\partial \beta_\lambda}{\partial \lambda}=\epsilon \,.
\ea

\section{Search strategy and results}\label{sec3}

In this section we study the system of $N$ fields at criticality using the one loop RG flow equations in the $\epsilon$-expansion below the upper critical dimension  $D_c=4$.
First we consider the case $N=2$ and perform a systematic search of fixed points of the RG flow giving all possible solutions.
Then we perform a partial analysis for the case $N=3$ and present here all the solutions which are charaterized by up to $5$ independent monomial composite operators in the critical potential.
Finally, observing the emergent pattern of symmetries, we then consider and study two infinite families of critical theories for any number $N$ of field flavors.

In general we show only the interacting solutions which are truly new and not {\it factorizable} in known solutions of a smaller number of fields. Since each solution is actually an orbit, given the underlining $O(N)$ symmetry which we require for the quadratic part of the effective action, i.e. for the two point function, we present them with the simplest representative we can find.

To distinguish among physically inequivalent fixed point solutions, which belong to different orbits, an important guide is given by the study of the invariant quantities under $O(N)$ transformations. In particular we have considered the anomalous dimension and the set of critical exponents, which are the eigenvalues of the stability matrix (gradient of the beta function vectors). 
As a further criterion we have considered only solutions for which exists a representative with couplings that can be or purely real or purely imaginary but not fully complex.
We stress that since the perturbative analysis is restricted to one loop, higher loop investigations could eventually disantangle possible degeracies
present in fixed points found at the lowest one loop order.

\subsection{Scale invariant solutions for $N=2$}

The case $N=2$ which has the six different couplings
\be
\!\!\!\!\! (\lambda_1,\,\,\lambda_2,\,\,\lambda_3,\,\,\lambda_4,\,\,\lambda_5,\,\,\lambda_6)=
(\lambda_{1,11},\,\, \lambda_{1,22},\,\, \lambda_{2,11},\,\, \lambda_{2,22},\,\, \lambda_{1,21}=
\lambda_{1,12},\,\, \lambda_{2,21}=\lambda_{2,12}) \,.
\label{couplings2}
\ee
Here we are interested in fixed points where possibly only one, two, three, four, or five couplings are different from zero, since by an $O(2)$ rotation at least one coupling can be set to zero.
Using the couplings in Eq.~\eqref{couplings2} the critical potential in general reads
\ba
\!\!\!\!\!\!\!\!V=\frac{i}{2} \Biggl(&{}& \lambda_1 \psi_{1}^{\dag} (\psi_{1}^{\dag}+\psi_{1})\psi_{1}+ \lambda_4 \psi_{2}^{\dag} (\psi_{2}^{\dag}+\psi_{2})\psi_{2}+  \lambda_2 (\psi_{2}^{\dag}{}^2 \psi_{1}+\psi_{1}^{\dag} \psi_{2}{}^2 ) +
\nonumber \\  &{}& \lambda_3 (\psi_{1}^{\dag}{}^2 \psi_{2}+ \psi_{2}^{\dag} \psi_{1}{}^2 )+ 
2 \lambda_{5} (\psi_{1}^{\dag} \psi_{1} \psi_{2}+\psi_{1}^{\dag}\psi_{2}^{\dag}\psi_{1})+2 \lambda_{6} ( \psi_{2}^{\dag} \psi_{1} \psi_{2}+\psi_{1}^{\dag}\psi_{2}^{\dag}\psi_{2} )
\Biggr)
\ea
so that the exchange of the two field flavors $(1,2)$ corresponds to the exchanges $(\lambda_1,\lambda_4)$, $(\lambda_2,\lambda_3)$ and $(\lambda_5,\lambda_6)$.

First of all we have imposed the conditions ${\cal C}$~\eqref{conditions}  of having counterterms to the two point function proportional to the unity $2 \times 2$ matrix, which amount to two extra equations, to be considered together with the six fixed point conditions.

We present the solutions found in Table~1 and in particular give for each critical theory a representative of the fixed point potential, its symmetry, and the anomalous dimensions $\gamma$ and $\tilde{\zeta}$. Moreover for each fixed point we have also computed the eigenvalues $e_n$ of the stability matrix (related to the six cubic operators which can be present in the potential).
Positive eigenvalues indicate which are the IR stable eigen-directions in the theory space charted by the couplings while UV stable directions correspond to the negative values. 
Moreover the solutions presented here (and in Table~2 as well) have $\nu_\perp=\frac{1}{2}+\frac{\kappa_\perp}{4}=\frac{1}{2}+\frac{3}{4} \tilde{\zeta}$.

We remind that the form of the potential of the solutions can change without affecting the physics thanks to the freedom of performing an $SO(2) (\subset O(2) )$ rotation, since the kinetic terms are invariant. Such a rotation has one generator and can be used to set to zero one coupling to simplify the searching procedure and picking up possibly a simpler representative of the orbit associated to a scaling solution.

Therefore we show a convenient representative for each different solution. Anomalous dimensions and eigenvalues of the stability matrix are clearly invariant under $O(2)$ field transformations.

We have included in the first row of the table also the "trivial" decoupled system of two self interacting single field (pomeron) solutions to show the stability properties. In the following we shall not show any other solution which can be factorized in terms of solutions with a smaller number of fields. Clearly such a solution as a $\mathbb{Z}_2$ symmetry for the exchange of the two fields and a 
$\mathbb{Z}_{PT}$ for the change of sign of the fields is combined with complex conjugation. 

Then we have found four new non trivial scaling solution.
The solution in the second row has a symmetry associated to the change of sign of $\psi_2$ combined with complex conjugation,  again named $\mathbb{Z}_{PT}$ with the two flavors interacting among each other. It has the same anomalous dimension but different stability properties compared to the first solution. The solution in the third row has the same anomalous dimensions of the previous two, other stability properties and the symmetry is a bit enhanced, since, apart  a $\mathbb{Z}_{PT}$ symmetry realized as before, there is also 
a $\mathbb{Z}_{2}$ component associated to the change of sign of $\psi_1$.

In the fourth and fifth rows of Table~1 we find two solution with different anomalous dimensions and new kind of symmetries.
In the fourth row the solution has a full diehedral point group symmetry $\mathbb{D}_3= S_3 $ which we shall see that lies at the bottom of a tower of solutions with permutational symmetry. Finally another solution with a cyclic symmetry has been found and report in the fifth row, which is also the first element of a family of solutions with more fields to be discussed later.

\begin{table}[]
\label{tablesolN2}
\begin{center}
\caption{Scale invariant solutions given by a representative of the cubic potential of each orbit, together with its finite symmetry encoded in the finite subgroup $G \subset O(2)$, the eigenvalues of the stability matrix and the anomalous dimensions $\tilde{\zeta}$ and $\gamma$. }
\tiny
\begin{tabular}{|l|l|l|l|l|l|}
\hline
 
&$V \frac{4 \pi }{\sqrt{\epsilon}}$& $G \subset O(2)$ & $(e_1,e_2,e_3,e_4,e_5,e_6)$ &$\tilde{\zeta} /\epsilon$ 
&$ \gamma /\epsilon  $  
 \\ \hline

1 & $\frac{i}{\sqrt{6}} \left[ \psi_1^\dagger \psi_1 (\psi_1+\psi_1^\dagger)  + \psi_2^\dagger \psi_2 (\psi_2+\psi_2^\dagger) \right] $
& $\mathbb{Z}_2 \times \mathbb{Z}_{PT}$  & $\frac{1}{6}(-2,-2,-1,0,5,6)$ & $\frac{1}{12} $& $-\frac{1}{12}$ \\ \cline{1-4}

2 &  \begin{tabular}{@{}c@{}}  $\frac{i}{\sqrt{6}}  \psi_1^\dagger \psi_1 (\psi_1+\psi_1^\dagger)  + i \sqrt{\frac{2}{3}} \psi_1^\dagger \psi_1 (\psi_2+\psi_2^\dagger) $ 
\\  $ +\frac{i}{\sqrt{6}}  \psi_2^\dagger \psi_2 (\psi_2+\psi_2^\dagger)  $ 
\end{tabular}
 & $\mathbb{Z}_{PT}$ & $\frac{1}{6}(-1,0,0,2,5,6)$  &  &  \\ \cline{1-4}

3 &  \begin{tabular}{@{}c@{}} 
$\frac{i}{2\sqrt{3}} \Bigl[  \psi_2^\dagger \psi_2 (\psi_2+\psi_2^\dagger) +2 \psi_1^\dagger \psi_1 (\psi_2+\psi_2^\dagger) $\\
$- (\psi_2^\dagger \psi_1^2+\psi_1^\dagger{}^2 \psi_2 )\Bigr]$
\end{tabular}
 & $\mathbb{Z}_2 \times \mathbb{Z}_{PT}$ & $\frac{1}{6}(-2,-2,-1,-1,0,6)$ &  &  \\   \hline
 
4 &  \begin{tabular}{@{}c@{}} 
$-\frac{1}{2} \psi_2^\dagger \psi_2 (\psi_2+\psi_2^\dagger) +\psi_1^\dagger \psi_1 (\psi_2+\psi_2^\dagger) $\\
$+\frac{1}{2} (\psi_2^\dagger \psi_1^2+\psi_1^\dagger{}^2 \psi_2 )$
\end{tabular} 
&  $\mathbb{D}_3= S_3  $ & $\frac{1}{2}(-7,-7,0,2,2,2)$   &  $-\frac{1}{4}$ & $\frac{1}{4}$ \\ \cline{1-4}

5 & $\frac{1}{\sqrt{2}} \left[ (\psi_2^\dagger \psi_1^2+\psi_1^\dagger{}^2 \psi_2 ) +
(\psi_1^\dagger \psi_2^2+\psi_2^\dagger{}^2 \psi_1 )\right]$
& $\mathbb{Z}_2$ & $\frac{1}{2}(-7, -2, -1, 0, 2, 2)$ &  &  \\ \hline

\end{tabular}
\end{center}
\end{table}

\subsubsection{Invariants}\label{Invariants}
In the following, even if we do not really need here, we briefly remind how one may construct explicit irreducible representations of $SO(2)$ under which transforms the potential in the basis of the couplings~\eqref{couplings2}. 
Considering an infinitesimal rotation of the fields in $\theta$ 
\be
\psi'=R \psi \, , \quad \psi^\dagger{}'=R \psi^\dagger \, , \quad
R=\left(
\begin{array}{cc}
 1 & \theta  \\
 -\theta  & 1 \\
\end{array}
\right) \,,
\ee
one finds the induced transformation on the coupling vector (potential)
\be
\tilde{R}_\theta^T =\left(
\begin{array}{cccccc}
 1 & 0 & \theta  & 0 & \theta  & 0 \\
 0 & 1 & 0 & \theta  & -\theta  & 0 \\
 -\theta  & 0 & 1 & 0 & 0 & \theta  \\
 0 & -\theta  & 0 & 1 & 0 & -\theta  \\
 -2 \theta  & 2 \theta  & 0 & 0 & 1 & \theta  \\
 0 & 0 & -2 \theta  & 2 \theta  & -\theta  & 1 \\
\end{array}
\right)  ,
\ee
which has eigenvalues $\rho_j$ and the corresponding eigenvectors $u_j$ in the complex form given by
\be
\left(
\begin{array}{ccl}
 \rho_1=1-i \theta \,\,  & u_1^T=& (0,-1,i,0,-i,1) \\
 \rho_2=1-i \theta  \,\,& u_2^T=&(i,i,1,1,0,0) \\
  \rho_3=1-3 i \theta \,\, & u_3^T=&\left(-\frac{1}{2},\frac{1}{2},\frac{i}{2},-\frac{i}{2},i,1\right) \\
 \rho_1^*=1+i \theta  \,\,& u_1^{*T}=&(0,-1,-i,0,i,1) \\
 \rho_2^*=1+i \theta \,\, & u_2^{*T}=&(-i,-i,1,1,0,0) \\
 \rho_3^*=1+3 i \theta \,\, & u_3^{*T}=&\left(-\frac{1}{2},\frac{1}{2},-\frac{i}{2},\frac{i}{2},-i,1\right) \\
\end{array}
\right)\,.
\ee
With these eigenvectors we can construct the combination of couplings  which transforms diagonally under an infinitesimal $SO(2)$  rotations as $u_i^T \lambda \to \rho_i u_i^T \lambda$. Note that the eigenvalues for a finite transformation are essentially phase factors.
Therefore monomials constructed as a product of powers the $u_i^T\lambda$ and $u_i^{*T}\lambda$, for which the product of the corresponding phase factor in the transformations becomes unity, are invariants. In this case it is evident that we can construct easily both quadratic and quartic invariants.
The anomalous dimensions and the eigenvalues of the stability matrix can therefore be written in terms of an independent set of the invariants.
\subsection{Scale invariant solutions for $N=3$}

We perform here a partial investigation for the three field case ($N=3$).
In this case the most generic potential is characterized by $18$ couplings. The conditions ${\cal C}$ of having counterterms from the two point function proportional to the unity $3 \times 3$ matrix, amount to five extra equations.  The $O(3)$ symmetry group contains the continuous $SO(3)$ group which has three generators,  and can be used to constrain three couplings. But in the following we are not attempting to perform an exhaustive search, for example generalizing what  was done in~\cite{Codello:2019isr,Codello:2020lta}  and we shall ignore this  possibility. Our strategy will consider the search of solutions characterized by few non zero couplings, most likely the simpler ones which we expect to be the most symmetric, and will stop at some point, leaving for future investigations a possible complete analysis.

\begin{table}[]
\label{tablesolN3}
\begin{center}
\caption{Scale invariant solutions given by a representative of a cubic potential of each orbit, together with its finite symmetry encoded in the finite subgroup $G \subset O(2)$. The corresponding eigenvalues of the stability matrix  are given in Table~3.}
\tiny
\begin{tabular}{|l|l|l|l|l|}
\hline
 
&$V \frac{4 \pi }{\sqrt{\epsilon}}$& $G \subset O(2)$ &$\tilde{\zeta}/\epsilon$ &$ \gamma /\epsilon  $   \\ \hline

1 & $\frac{1}{\sqrt{2}} \left(\psi_1 \psi^\dagger_2{}^2+\psi_2 \psi^\dagger_3{}^2+ \psi_3 \psi^\dagger_1{}^2\right) + (\psi_i \leftrightarrow \psi^\dagger_i)$
& $\mathbb{Z}_3$   & -$\frac{1}{4} $&  $\frac{1}{4}$ \\ \cline{1-5}

2 & $ i \left(\psi_1 \psi^\dagger_2 \psi^\dagger_3+\psi_2 \psi^\dagger_3\psi^\dagger_1+ \psi_3 \psi^\dagger_1 \psi^\dagger_2\right) + (\psi_i \leftrightarrow \psi^\dagger_i)$
& $S_4 \times \mathbb{Z}_{PT}$   & $\frac{1}{4} $&  -$\frac{1}{4}$ \\ \cline{1-3}

3 & $ i \left(\psi_1 \psi^\dagger_2 \psi^\dagger_3-\psi_2 \psi^\dagger_3\psi^\dagger_1+ \psi_3 \psi^\dagger_1 \psi^\dagger_2\right) + (\psi_i \leftrightarrow \psi^\dagger_i)$
& $ \mathbb{Z}_2^3 \times \mathbb{Z}_{PT}$   &  & \\ \cline{1-3}

4 & $ i \left( \psi_1 \psi^\dagger_2 \psi^\dagger_3+\psi_2 \psi^\dagger_1\psi^\dagger_3 + \frac{1}{2} (\psi_1^2+\psi_2^2)\psi_3^\dagger )\right)+ (\psi_i \leftrightarrow \psi^\dagger_i)$
& $ \mathbb{Z}_2^2 \times \mathbb{Z}_{PT}$   &  & \\ \hline

5 & $ \frac{i}{\sqrt{3}} \psi^\dagger_1 \psi_1 \psi_3+\frac{i}{2\sqrt{3}} \psi^\dagger_3 (  \psi_3^2 -  \psi_1^2 )
+ \frac{i}{\sqrt{2}} \psi^\dagger_2 \psi_2^2 - \sqrt{\frac{2}{3}} \psi_1 \psi_2 \psi^\dagger_2
+ (\psi_i \leftrightarrow \psi^\dagger_i)$
& $ \mathbb{Z}_{PT} $   &$\frac{1}{12}$ & $-\frac{1}{12} $\\ \cline{1-3}

6 & $ \frac{i}{2} \sqrt{\frac{3}{2}} \psi_3 \left( \psi^\dagger_1 \psi_1+\psi^\dagger_2 \psi_2 \right)+ \frac{i}{\sqrt{6}} \psi_3 \psi^\dagger_3{}^2+ \frac{1}{2\sqrt{6} }\psi_3 \left( \psi_1\psi^\dagger_2-\psi_2\psi^\dagger_1\right)
+ (\psi_i \leftrightarrow \psi^\dagger_i)$
& $ \mathbb{Z}_{8} $    &  & \\ \cline{1-3}

7 & $i \sqrt{\frac{2}{3}} \psi_3 \left( \psi^\dagger_1 \psi_1+\psi^\dagger_2 \psi_2 \right)+ \frac{i}{\sqrt{6}} \psi_3 \psi^\dagger_3{}^2-
\frac{1}{\sqrt{6} }\left( \psi_1\psi^\dagger_1{}^2+\psi_2\psi^\dagger_2{}^2\right)
+ (\psi_i \leftrightarrow \psi^\dagger_i)$
& $  \mathbb{Z}_{2} \times  \mathbb{Z}_{PT} $   &  & \\ \cline{1-3}

8 & $ \frac{i}{2\sqrt{3}}  \psi_3 \left( \psi^\dagger_3{}^2 - \psi^\dagger_2{}^2 \right)+ 
i \psi_3 \left( \frac{\sqrt{3}} {2} \psi_1 \psi^\dagger_1+\frac{1}{\sqrt{3}} \psi_2 \psi^\dagger_2 \right) -\frac{1}{2}\sqrt{\frac{5}{3}}\psi_1\psi^\dagger_1\psi_2
+ (\psi_i \leftrightarrow \psi^\dagger_i) $
& $  \mathbb{Z}_{2} \times \mathbb{Z}_{PT} $    &  & \\ \hline

\end{tabular}
\end{center}
\end{table}
We present them in Table~2 and give the eigenvalues of the stability matrix in Table~3.
We do not show the solution with three equal decoupled pomerons, which is clearly present. It has $3$ positive, $3$ null and $12$ negative eigenvalues of the stability matrix. 
We do not show also all the solutions characterized by one reggeon system decoupled from  a coupled system of the other two, since they can be constructed by combining  a single pomeron system solution with the solutions given in the previous subsection. The only piece of information which could be of interest is the one regarding the stability properties looking at the eigenvalues of the stability matrix, which can be easily obtained in case of need.

We consider and present here all the non trivial solutions (independent orbits) which have up to five non zero couplings.

With three non zero couplings we find three solutions, which in Table~2 are numbered from $1$ to $3$.
The first one has a real interacting potential, cyclic symmetric in the flavor, corresponding to a $\mathbb{Z}_3$ symmetry. We note that a solution with such symmetry property has been discussed in~\cite{Janssen:2000ds}.
It is the second member of a general family of cyclic symmetric scaling solutions for any $N$ which will be discussed later.
The second solution has an $S_4 \times \mathbb{Z}_{PT}$ , where  with $\mathbb{Z}_{PT}$ we means a change of sign and complex conjugation. Also this is the second instance of a solution belonging to a general family, this time with $S_{N+1}$ permutation symmetry. They will be discussed in general in the next subsection. 
The third solution in Table~2 is the last non trivial one that we have found allowing only three non zero couplings. It looks similar to the previous one apart from a sign changed which makes it is less symmetric, specifically with a $\mathbb{Z}_2^3 \times \mathbb{Z}_{PT}$  symmetry. The $3$ factors $\mathbb{Z}_2$  denote the symmetry under the exchange of the fields $(1,3)$, the change of sign of both of them and the change of sign of another pair of fields, e.g. $(1,2)$.

Allowing a fourth non zero coupling we find another non trivial solution (number $4$ in Table~2) with symmetry  $\mathbb{Z}_2^2 \times \mathbb{Z}_{PT}$, realized by exchange of the fields $(1,2)$ and changing the sign of both of them. As usual the rest of the symmetry present is related to a change of sign and complex conjugation.

Finally allowing five non zero couplings we find four more solutions numbered from $5$ to $8$ in Table~2.
The first (number $5$) has just a $\mathbb{Z}_{PT}$ symmetry realized by changing sign of the pair of fields $(2,3)$ and performing a complex conjugation. The solution number $6$ has a cyclic $\mathbb{Z}_8$ symmetry 
describing the invariance under rotation with an angle $\pi/4$  in the plane of the field flavors $(1,2)$. 
The solution in the $7$th row of Table~2 is symmetric under exchange of fields $(1,2)$ and under change of sign of field $3$ together with complex conjugation. The last solution (number $8$) has a $\mathbb{Z}_2$  factor because it  is symmetric under the change of sign of field $1$ and the same realization of $\mathbb{Z}_{PT}$ as in the previous solution.

The degree of stability under perturbations according to the RG flow of all these $8$ fixed point solutions is reported in Table~3.

We have also analyzed the case with $6$ non zero couplings finding other independent solutions 
but we do not find it worth to report them here.

\vskip 1 cm
\begin{table}[]
\label{tablesolN3stability}
\begin{center}
\caption{Eigenvalues of the stability matrix  of the solutions presented in Table~2. For the solution 1 and 5 we give the numerical values when the analytic form in constructed from roots of quartic equations.}
\tiny
\begin{tabular}{|l|l|l|l|l|l|}
\hline
 
&\hspace{6.5cm}$(e_1,\cdots ,e_{18})$    \\ \hline

1 & \begin{tabular}{@{}c@{}}  $ (-1.76181,-1.76181,-1.76181,-0.720398-0.832342 i,-0.720398-0.832342 i,-0.720398-0.832342 i,-0.720398+0.832342 i, $ \\
$ 0.720398+0.832342   i,-0.720398+0.832342 i,0.,0.,0.,0.\, -0.866025 i,0.\, +0.866025 i,0.702608,0.702608,0.702608,1.) $ 
\end{tabular} \\ \hline

2 & $ (-\sqrt{7},-\sqrt{7},-\sqrt{7},-2,-2,-2,-1,-1,-1,-1,-1,0,0,0,1,\sqrt{7},\sqrt{7},\sqrt{7})$  \\ \hline

3 & $ \left(-\sqrt{7},-\frac{1}{2} (3+\sqrt{5}),-\frac{1}{2} (3+\sqrt{5}),-2,-1,-1,-1,-1,-1,\frac{1}{2}
   (\sqrt{5}-3),\frac{1}{2} (\sqrt{5}-3),0,0,0,0,0,1,\sqrt{7} \right) $  \\ \hline
   
4 & $    \left(-\sqrt{7},-\sqrt{7},-\frac{1}{2} (3+\sqrt{5}),-\frac{1}{2} (3+\sqrt{5}),-2,-1,-1,-1,-1,\frac{1}{2}
   \left(\sqrt{5}-3\right),\frac{1}{2} \left(\sqrt{5}-3\right),0,0,0,0,1,\sqrt{7},\sqrt{7}\right) $  \\ \hline

5  &  \begin{tabular}{@{}c@{}}   $\bigl( \, -\frac{1}{3} (1+\sqrt{2}),-\frac{2}{3},-0.603553-0.441845 i,-0.603553+0.441845 i,-0.383164,-\frac{1}{3},-\frac{1}{3}, $ \\ 
$-\frac{1}{6},-\frac{1}{6},0,0,0,0,\frac{1}{3} (\sqrt{2}-1),0.590271,1,\frac{1}{12} (15-i \sqrt{15}),\frac{1}{12} (15+i \sqrt{15}) \,\bigr)$
\end{tabular} \\ \hline

6  & $\left( -\frac{1}{6},0,0,0,0,0,0,0,\frac{1}{4},\frac{1}{4},\frac{1}{4},\frac{1}{3},\frac{1}{3},\frac{1}{2},\frac{1}{2},\frac{5}{6},\frac{5}{
   6},1\right)$ \\ \hline

7  & $\left( -\frac{1}{6},-\frac{1}{6},0,0,0,0,0,\frac{1}{3},\frac{1}{3},\frac{1}{3},\frac{1}{3},\frac{1}{3},\frac{1}{3},\frac{2}{3},\frac{5}{6},
   \frac{5}{6},\frac{5}{6},1\right)$ \\ \hline
   
8  & $\left( -\frac{1}{3},-\frac{1}{3},\frac{1}{6} (1-\sqrt{5}),-\frac{1}{6},-\frac{1}{6},-\frac{1}{6},\frac{1}{12}
   (1-\sqrt{5},0,0,0,0,0,\frac{1}{12} (1+\sqrt{5}),\frac{1}{6} (4-\sqrt{5}),\frac{1}{2},\frac{1}{6}
   (1+\sqrt{5}),1,\frac{1}{6} (4+\sqrt{5} )\right)$ \\ \hline

\end{tabular}
\end{center}
\end{table}

\subsection{A family of  $S_{N+1}$ symmetric fixed point solutions}\label{SNplus1sol}
Having observed that for $N=2$ exists a fixed point solution ($4$th row in Table~1) with $S_3$ symmetry and for $N=3$ one with $S_4$ symmetry, it is natural to expect the existence for $N$ fields of a fixed point invariant under point group $S_{N+1}$.

For this highly symmetric family of solutions the interacting potential can be described by a single coupling according to
\be
\lambda_{i,jk}= \lambda \,q^{(3)}_{ijk}
\ee
where the fully symmetric $q^{(3)}$ tensor is defined and discussed in Appendix~\ref{qtensors}. Let us stress that this representation is valid for all $N\ge2$, while the case $N=1$ is not included in this description.

As previously discussed we remind that the explicit form of the potential of these critical theories is not uniquely determined as we know that it is associated to a full orbit of physically equivalent fixed point potentials transforming under a representation of the coset $O(N)/S_{N\!+\!1}$.
For example let us consider the two solutions with this symmetry previously found for $N=2$ and $N=3$.
The case $N=2$ can be obtained considering the following set of three vectors of  ${\mathbb R}^2$
\be
e^1=\sqrt{2} (0, -1)^T \,, \,e^2=\frac{\sqrt{2}}{2} (\sqrt{3}, 1)^T \,,\, e^3=\frac{\sqrt{2}}{2} (-\sqrt{3}, 1)^T
\ee
which correspond to the vertices of an equilateral triangle,
while for the case $N=3$ the explicit solution found correspond to the set of the four vectors of  ${\mathbb R}^3$
\be
e^1=(1, 1, -1)^T \,, \, e^2=(-1, -1, -1)^T \,, e^3=(1, -1, 1)^T\,, \, e^4=(-1, 1, 1)^T
\ee
which correspond to the vertices of a tetrahedron.

Using the relation~\eqref{qq} the counterterms for the two point function becomes diagonal in the flavor and automatically satisfy the constraints in Eq.~\eqref{conditions}. In particular we obtain
\be
\gamma=-\frac{N-1}{8} \frac{\lambda^2}{(4\pi)^2} \,, \quad \tilde{\zeta}=\frac{N-1}{8} \frac{\lambda^2}{(4\pi)^2} \,, \quad 
\kappa_\perp=\frac{3}{8} (N-1) \frac{\lambda^2}{(4\pi)^2} \,.
\ee

The counterterm associated to the triple interation is strongly simplified thanks to the relation~\eqref{qqq}. 
Putting all the results together as in Appendix~\ref{oneloopcomp} one obtains a simple form
for the beta function for the coupling $\lambda$. It is convenient to introduce the rescaled coupling $\tilde{\lambda}=\lambda/(4\pi)$,
 for which the beta function reads
\be
\beta_{\tilde{\lambda} }= -\frac{\epsilon}{2} \tilde{\lambda} -\frac{N\!-\!1}{4}  \tilde{\lambda}^3 +(N\!-\!2) \tilde{\lambda}^3 \,.
\ee

We note that for $N=2$ the vertex correction at one loop is zero and the fixed point is non trivial because of a non zero anomalous dimension. 
At the fixed point one has
\be
 \tilde{\lambda}_*^2=\frac{2\epsilon}{3N\!-\!7} \,, \quad -\gamma=  \tilde{\zeta}=\kappa_\perp/3=
 \frac{N\!-\!1}{3N\!-\!7} \frac{\epsilon}{4} \, , \quad 
 \frac{\partial \beta_{\tilde{\lambda} }} {\partial \tilde{\lambda} }=\epsilon\,.
\ee

One can easily check that for example for $N=2,3,4$ one has $\gamma= \frac{\epsilon}{4}, -\frac{\epsilon}{4}, -\frac{3}{20} \epsilon$ so that the previously obtained results for $N=2$ and $N=3$ are reproduced.
One obtains also
\be
\nu_\perp=\frac{1}{2} +\frac{3}{16} \frac{N\!-\!1}{3N\!-\!7}\epsilon \,.
\ee

For $N\ge 3$ the fixed point coupling $\lambda_*$ is real and therefore the symmetry is enhanced to include a global change of sign of the fields together with a complex conjugation, which we name with $\mathbb{Z}_{PT}$.
Curiously, one can notice that the properties associated to $N=1$,  single pomeron case, are reproduced in the limit $N \to \infty$.

It is well known that an Euclidean theory of a scalar field multiplet with a symmetric cubic interaction as above is describing a Potts model and there exist meaningful cases for the limit  $N\to0$ and $N\to-1$, associated to percolation and spanning forest statistical models respectively.
It is not clear if in our nonisotropic case such limits can have some kind of physical interpretation.

\subsection{A family of  $\mathbb{Z}_N$ symmetric fixed point solutions}\label{ZNsol}

Solutions with a cyclic symmetry were found for  $N=2$ ($5$th row in Table~1) and $N=3$ ($1$st row in Table~2).
It is natural to expect in general the existence of a cyclic symmetric ($\mathbb{Z}_N$) scale invariant solution, 
and so we test here this idea. A simple realization of a cyclic symmetric potential is given by 
\be
V_{\rm cyclic}= c \left[ \psi_1 \psi^\dagger_2{}^2+\psi_2 \psi^\dagger_3{}^2+\cdots + \psi_N \psi^\dagger_1{}^2 + (\psi_i \leftrightarrow \psi^\dagger_i) \right] \,,
\ee
a form which can be obtained inserting in~\eqref{action} the coupling 
\be
\lambda_{i,jk}=\lambda\, \delta_{jk} \delta_{j,1+i \,{\rm mod} \,N} \,.
\ee
From this definition we find for the coupling dependence in the one loop two- and three- point functions (and therefore in the counterterms) that
\be
\lambda_{i,ml}\lambda_{j,ml} = \lambda^2 \delta_{ij} \,, \quad  \lambda_{i,ab}  \lambda_{j,ac} \lambda_{b,ck} = 0 \,.
\ee
Using these results and having again defined $\tilde{\lambda}=\lambda/(4\pi)$, we find
\be
-\gamma=  \tilde{\zeta}=\kappa_\perp/3= \frac{\tilde{\lambda}^2}{8}
\ee
and, with the vertex correction being zero, the beta function acquires a very simple form
\be
\beta_{\tilde{\lambda} }= -\frac{\epsilon}{2} \tilde{\lambda} +\tilde{\zeta} \lambda+3\gamma \lambda
=-\frac{\epsilon}{2} \tilde{\lambda} - \frac{\tilde{\lambda}^3}{4} \,,
\ee
giving at the fixed point, independently from $N$~\footnote{This is in agreement with a statement in~\cite{Janssen:2000ds}.},

\be
\tilde{\lambda}_*^2 = -2\epsilon \,, \quad \gamma= - \tilde{\zeta}=- \kappa_\perp/3=\frac{\epsilon}{4} \, , \quad 
\nu_\perp=\frac{1}{2} -\frac{3}{16}\epsilon \, , \quad 
 \frac{\partial \beta_{\tilde{\lambda} }} {\partial \tilde{\lambda} }=\epsilon\,.
\ee
This is a phenomenon similar to what has been seen to happen in the first not trivial order of perturbation theory of cubic higher derivative shift symmetric theories~\cite{Safari:2021ocb}.

\section{Conclusions}
We have explored specific realizations of multi-field RFT having the Reggeon fields with the same Regge trajectory, 
preserved by the interactions, which are assumed a priori arbitrary. The analysis of the scale invariant critical solutions has been performed with renormalization group techniques in the first non trivial order in the $\epsilon$-expansion below the upper critical dimension.

Our investigations have covered the case of $N=2$ and $N=3$ fields (non exautive for the latter) and revealed the pattern of critical theories with their global symmetries. The zoology of solutions grows rapidly with $N$ and even if it is possible to push the search further, it would reach soon the possibilities of an analytical search and would require heavy numerical investigations. 
We have prolonged solutions found for the two and three field cases to construct two scale invariant models for an arbitrary $N$.

Given the equivalence among RFT and out-of-equilibrium Directed Percolation models, this analysis can be of interest in multi-field extensions of these statistical models. These could be interesting since the role of symmetries could help in the understanding of non equilibrium dynamics close to criticality.

We have chosen to impose on the system the conditions of having a single Regge trajectory to see in theories as simple as possible the pattern of symmetries compatible with scale inviariance, but certainly it will be interesting and deserves future investigations the case where such apriori conditions are removed. In this way all possible scaling solutions can be investigated, 
taking into account the possible residual $O(m<N)$ symmetry of any RG fixed point if $m<N$ Regge trajectories happen to coincide.

A natural step to improve the current understanding is to push the investigation beyond perturbation theory.
Apart from numerical montecarlo studies the alternative theoretical approach is to perform a Wilsonian RG analysis using for example functional renormalization group methods along the line previoulsy employed in the study of the pomeron odderon system~\cite{Bartels:2016ecw}. We hope to address this problem in a near future but of course in the multifield case this is a computationally hard task in its generality, and possibly easier fixing the specific global symmetries to be investigated.
One has also to keep   in mind that given that in  this theory a criticality is not a CFT (and is also not unitary) there is no possibility to use conformal bootstrap~\cite{Rattazzi:2008pe} methods.

\section{Acknowledgement}
This research was supported by the Fondecyt Fund (Chile)  grands N:1231829. J.B. thanks the support of the International Collaboration Fondecyt F:1231829/Chile and
also the hospitality of the Federico Santa María Technical University. C.C also appreciates the support of the INFN
and the hospitality of the University of Bologna and the Hamburg University.
\appendix

\section{One-loop beta functions and anomalous dimensions} \label{oneloopcomp}

We study here the one loop renormalization properites of multi-field Reggeon Field Theory (equivalent to the multi-field Directed Percolation) focusing on a special highly symmetric case with just $O(N)$ invariant quadratic operators. The theory is defined in terms of $N$ reggeon fields $\psi_i$ and $\psi_i^\dagger$ (order parameters) on a $D+1$ dimensional domain, constructed as an euclidean $D$-dimensional transverse space with coordinate $x$ times a one dimensional time space with coordinate $\tau$. The dual variables will be momentum $q$ with mass dimension $M$ and energy $E$.

The bare action
\ba
S&=&\int d\tau d^D x \Big[
\psi_{B}^{\dag}{}^{T} \left( i\partial{\tau} + \alpha'_B \nabla^2 + \tilde{\alpha}'_B \mu_B \right) \psi_{B}]\nonumber\\
&-&\frac{i}{2} \tilde{\alpha}'_B  \lambda_{B\,i,jk} \left(\psi_{B;i}^{\dag}\psi_{B;j} \psi_{B;k}+\psi_{B,i} \psi_{B,j}^{\dag} \psi_{B;k}^{\dag} \right)
 \Big] 
\ea
is written in terms of the bare quantities.
Here $\psi_i$ ($\psi_i^\dagger$) is an element of the $N$-component field vector $\psi$ ($\psi^\dagger$), "$T$" indicates transposition. In the most general case $\alpha'$ and  $\mu$ are symmetric matrices and one of them can always be made diagonal ($\alpha'$) employing the $O(N)$ symmetry of the $\tau$ derivative term in the two-point function. 
We have introduced the scalar quantity  $\tilde{\alpha}'$ in flavor space (number) constructed from $\alpha'$, which is needed for dimensional reasons and defines the different anomalous scaling in the transverse space with respect to the time space.

In this work we shall consider the special restricted case with such matrices {\it proportional to the identity}, so that $\tilde{\alpha}'$  and 
$\alpha'$ in practice coincide with a trivial abuse of notation, making bare slopes and masses independent of the flavor $i$.  
This translates into a requirement on the form of the counterterms of the two-point function which, instead of being generically symmetric, have to be proportional to the identity matrix (i.e. one has to impose $N(N+1)/2-1$ conditions on the couplings at the fixed point).
We shall also specialize to the case $\mu=0$ which denotes scale invariance in dimensional regularization. 
The $\lambda_{i,jk}=  \lambda_{i,kj}$ are symmetric in $(j,k)$ and summation over the flavor indices is understood.

From dimensional analysis we have
\ba
[\tau]=E^{-1}\,, \quad [x]=M^{-1} \,, \quad [\psi_B]=M^{D/2} \,, \quad [\alpha']=E/M^2 \,, \quad [\mu]=M^2 \,, \quad [\lambda_{i,jk}]=M^{2-D/2} 
\nonumber
\ea
from which one can see that the critical dimension for a dimensionless coupling associated to the cubic interaction is $D_c=4$.

The renormalized form of the action, for $D=D_c-\epsilon$, is given by
\ba
S&=&\int d\tau d^D x \Big[ \psi^{\dag}{}^{T} Z_{\psi} i \partial_{\tau} \psi +  {\psi^{\dag}}^{T} Z_{\psi}^{\frac{1}{2}}
Z_{\alpha'} \alpha' \nabla^2 Z_{\psi}^{\frac{1}{2}} \psi + 
 {\psi^{\dag}}^{T} Z_{\psi}^{\frac{1}{2}}
Z_{\mu} \tilde{\alpha}' \mu Z_{\psi}^{\frac{1}{2}} \psi \nonumber\\
&&-\frac{i}{2} \tilde{\alpha}' \left(\lambda_{i,jk} +{\delta\lambda}_{i,jk}\right) M^{\frac{\epsilon}{2}} \left(\psi_{i}^{\dag}\psi_{j} \psi_{k}+\psi_{i} \psi_{j}^{\dag} \psi_{k}^{\dag} \right)
\Big] \,,
\ea
where all the $Z'$s, constructed by suitable counterterms, are in general matrices while the vertex counterterm ${\delta\lambda}_{i,jk}$ is a three index tensor. 
The vector field is renormalized as
\ba
\psi_B=Z_{\psi}^{\frac{1}{2}} \psi \,, \quad Z_{\psi}=1+\delta_\psi \, 
\ea
with $Z_\psi$ a symmetric matrix constructed from the counterterm $\delta_\psi$.

Keeping this relation into account one can obtain the usual relations among bare and renormalized quantities.
We define the matrices
\be
Z_{\alpha'}\alpha'= \alpha_B' \, , \quad Z_{\mu} \tilde{\alpha}'  \mu= \tilde{\alpha}'_B \mu_B,
\ee
\be
Z_{\alpha'}=1 + \delta_{\alpha'}\, , \quad Z_{\mu}=1 + \delta_\mu\,.
\ee
As said we also define 
\be
\tilde{Z}_{\alpha'} \tilde{\alpha}'= \tilde{\alpha}_B' \, , \quad \tilde{Z}_{\alpha,}=1 + \tilde{\delta}_{\alpha'}  ,
\ee
where we choose $\tilde{\alpha}'$ according to
\ba
\tilde{Z}_{\alpha'}= \frac{1}{N} tr \left( Z_{\alpha'}\right)  = 1+\frac{1}{N} tr \left(\delta_{\alpha'} \right)\,.
\ea


For the couplings one has
\ba
\lambda_{B;i,jk}= M^{\frac{\epsilon}{2}} \tilde{Z}_{\alpha'}^{-1} \left( \lambda_{i',j'k'}+\delta \lambda_{i',j'k'} \right)(Z_{\psi}^{-\frac{1}{2}})_{i'i} (Z_{\psi}^{-\frac{1}{2}})_{j'j} (Z_{\psi}^{-\frac{1}{2}})_{k'k}  ,
\ea
and to derive the beta function we impose the usual condition  $M \frac{\partial}{\partial M} \lambda_{B;i,jk}=0$
which leads to 
\ba
0 &=&  \Big[ \frac{\epsilon}{2}   (\lambda +\delta  \lambda)_{i,jk} -M \frac{\partial}{\partial M}  \log \tilde{Z}_{\alpha'}(\lambda +\delta  \lambda)_{i,jk}+ M \frac{\partial}{\partial M}  (\lambda+ \delta \lambda)_{i,jk}  \nonumber\\
&&\hspace{1cm}   - \frac{1}{2}  \left(   (M \frac{\partial}{\partial M}    \log Z_{\psi} )_{i''i}   (\lambda+ \delta \lambda)_{i'',jk}     
+ (M \frac{\partial}{\partial M}    \log Z_{\psi} )_{j''j}   (\lambda+ \delta \lambda)_{i,j''k} \right.  \nonumber\\ 
&&
\left.+ (M \frac{\partial}{\partial M}    \log Z_{\psi} )_{k''k}   (\lambda+ \delta \lambda)_{i,jk''} \right) \Big]\,.
\ea
With the anomalous dimension matrix
\ba
\gamma= \frac{1}{2} M \frac{\partial}{\partial M} \log Z_{\psi}
\ea
and with
\ba
\tilde{\zeta}_{\alpha}=    M \frac{\partial}{\partial M}        \tilde{Z}_{\alpha}
\ea
and 
\ba
\beta_{i,jk}= M \frac{\partial}{\partial M} \lambda_{i,jk} =-\frac{\epsilon}{2} \lambda_{i,jk} + \tilde{\beta}_{i,jk}
\ea
and 
\ba
 M \frac{\partial}{\partial M}  \delta \lambda_{i,jk} = \beta_{l,mn} \frac{\partial}{\partial \lambda_{l,mn}} \delta \lambda_{i,jk}
 \approx  -\frac{\epsilon}{2} \lambda_{l,mn} \frac{\partial}{\partial \lambda_{l,mn}}\delta \lambda_{i,jk} \,,
\ea
we get
\ba
0&=&-\tilde{\zeta}_{\alpha} (\lambda +\delta  \lambda)_{i,jk} - \gamma_{i'i}(\lambda +\delta  \lambda)_{i',jk}-
\gamma_{j'j} (\lambda +\delta  \lambda)_{i,j'k}-\gamma_{k'k}(\lambda +\delta  \lambda)_{i,jk'}  \nonumber\\
&&+\tilde{\beta}_{i,jk} +\frac{\epsilon}{2} \delta \lambda_{i,jk} + \beta_{l,mn} \frac{\partial}{\partial \lambda_{l,mn}} \delta \lambda_{i,jk}
\ea
or
\ba
\label{betatilde}
\tilde{\beta}_{i,jk}&\approx  &\tilde{\zeta}_{\alpha} \lambda_{i,jk}+(\gamma_{i'i} \lambda_{i',jk}+
\gamma_{j'j} \lambda_{i,j'k}+\gamma_{k'k}\lambda_{i,jk'} )\nonumber\\
&& -\frac{\epsilon}{2} \delta \lambda_{i,jk}+ \frac{\epsilon}{2} \lambda_{l,mn}\frac{\partial}{\partial \lambda_{l,mn}} \delta \lambda_{i,jk} \,.
\ea

Let us now derive explicitely the counterterms computing the one loop integrals. 
We begin with the integral associated to the 2-point  diagram of Fig.1:
\begin{figure}[H]
\begin{center}
\epsfig{file=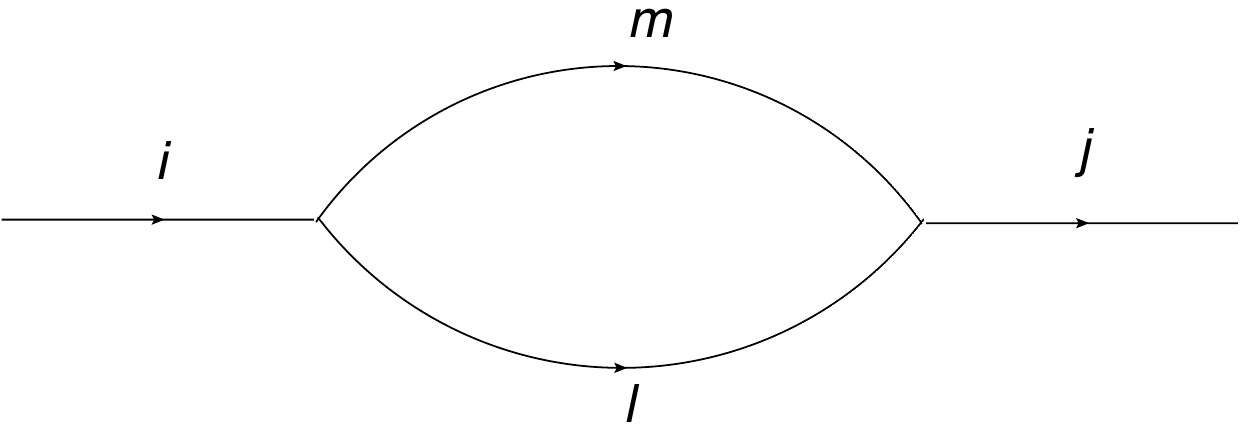,width=6cm,height=8cm}\\
\end{center}
\caption{One loop 2-point function} 
\label{Fig1}
\end{figure}
In leading order $\epsilon$ one obtains
\ba
\label{gamma11eps}
i(\Gamma^{(1,1)}_{kl;div})_{ij}&=& \frac{1}{4 \epsilon}   \frac{\lambda_{i,ml}\lambda_{j,ml}}{(4\pi)^2} 
 \Big[ E-\frac{\alpha'}{2} q^2 +2 \tilde{\alpha'}  \mu  \Big]  \,.
\ea
Clearly all the three counterterms are proportional to the same symmetric matrix.
At this point we impose here the conditions ${\cal C}$ to have all of them proportional to the identity matrix:
\ba
&& \lambda_{i,ml}\lambda_{j,ml} =0 \, , {\rm for} \, i \ne j \nonumber \\
&& \lambda_{1,ml}\lambda_{1,ml} = \lambda_{2,ml}\lambda_{2,ml}  \cdots = \lambda_{N,ml}\lambda_{N,ml}  ,
\label{conditions}
\ea
which imply that
\be
 \lambda_{i,ml}\lambda_{j,ml}  =\frac{1}{N}  \lambda_{k,ml}\lambda_{k,ml}   \delta_{ij}\,.
\ee

In general these one loop corrections then lead to:
\ba
Z_{\psi_{ij}}&=&1+ \delta \psi_{ij}= 1 + \frac{1}{\epsilon} \frac{1}{4(4\pi)^2} \lambda_{i,lm} \lambda_{j,lm} 
\ea
and 
\ba
\left( Z_{\psi}^{\frac{1}{2}} Z_{\alpha'}Z_{\psi}^{\frac{1}{2}} \right)_{ij}=1 + \frac{1}{\epsilon} \frac{1}{8(4\pi)^2} \lambda_{i,lm} \lambda_{j,lm}
\ea
which implies
\ba
Z_{\alpha'}=1-\frac{1}{\epsilon} \frac{1}{8(4\pi)^2} \lambda_{i,lm} \lambda_{j,lm}
\ea
and
\ba
\tilde{Z}_{\alpha'}&=&\frac{1}{N} tr(Z_{\alpha'})= 1-\frac{1}{\epsilon} \frac{1}{8(4\pi)^2} \frac{1}{N}\lambda_{i,lm} \lambda_{i,lm} \,.
\ea
The third term:
\ba
\left( Z_{\psi}^{\frac{1}{2}} Z_{\mu}Z_{\psi}^{\frac{1}{2}} \right)_{ij}=1 + \frac{1}{\epsilon} \frac{1}{2(4\pi)^2}\lambda_{i,lm} \lambda_{j,lm}  ,
\ea
which implies
\ba
 (Z_{\mu})_{ij}=1 + \frac{1}{\epsilon}  \frac{1}{4(4\pi)^2} \lambda_{i,lm} \lambda_{j,lm}
\ea
and
\ba
\left( \tilde{Z}_{\alpha'}^{-1} Z_{\mu} \right)_{ij}=1 + \frac{1}{\epsilon} \frac{3}{8}
 \frac{1}{(4\pi)^2}\lambda_{i,lm} \lambda_{j,lm} \,.
\ea
From this one can compute the anomalous dimensions.
In our specific case, because of the conditions~\eqref{conditions} all the fields have the same anomalous dimensions and there is also just one anomalous dimension for the $\psi^\dagger{}^T \psi$ O(N) symmetric composite operator related to the "mass" $\mu$.  

We can therefore obtain in a usual way the anomalous dimensions:
\ba
\gamma_{ij}&=& \frac{1}{2} \left( -\frac{\epsilon}{2} \lambda_{r,st} \right)  \frac{\partial}{\partial \lambda_{r,st}} \delta \psi_{ij}
 =-\frac{1}{8}  \frac{1}{(4\pi)^2}\lambda_{i,lm} \lambda_{j,lm} = \gamma\, \delta_{ij}\,,
\ea
where the last equality derives from the condition~\eqref{conditions}, and
\ba
\tilde{\zeta}  
&=&-\frac{1}{2N} \lambda_{r,st}  \frac{\partial}{\partial \lambda_{r,st}} \frac{-1}{8(4\pi)^2} \lambda_{i,lm} \lambda_{i,lm}
=\frac{1}{N}  \frac{1}{8(4\pi)^2}\lambda_{i,lm} \lambda_{i,lm}\,.
\ea
This model is characterized by a second order phase transition and approaching criticality there are two different correlation lengths,
one for the transverse space $\xi_\perp \sim |\mu|^{-\nu_\perp}$ and one for time (rapidity) $\xi_\parallel \sim |\mu|^{-\nu_\parallel}$, with the usual definition $z=\nu_\parallel/\nu_\perp$. At leading order one has $z=2-\tilde{\zeta}$.
Another useful quantity is
\be
{\kappa_\perp}_{ij}=- \left( -\frac{\epsilon}{2} \lambda_{r,st} \right)  \frac{\partial}{\partial \lambda_{r,st}} \ln{\left( \tilde{Z}_{\alpha'}^{-1} Z_{\mu} \right)}_{ij} =\frac{3}{8} \frac{1}{(4\pi)^2}\lambda_{i,lm} \lambda_{j,lm}
= \kappa_\perp \, \delta_{ij} \,,
\ee
from which one can compute at one loop~\cite{Bartels:2015gou} the critical exponent

\be
\nu_\perp= (2- \kappa_\perp)^{-1}\simeq \frac{1}{2}+\frac{1}{4} \kappa_\perp \,.
\ee
Clearly  $\gamma$, $\tilde{\zeta}$ and $\kappa_\perp$ are all proportional up to a numerical factor.

The counterterms for the triple interations can be computed analyzing the one loop contributions to the three point function in Fig.2:
\begin{figure}[H]
\begin{center}
\epsfig{file=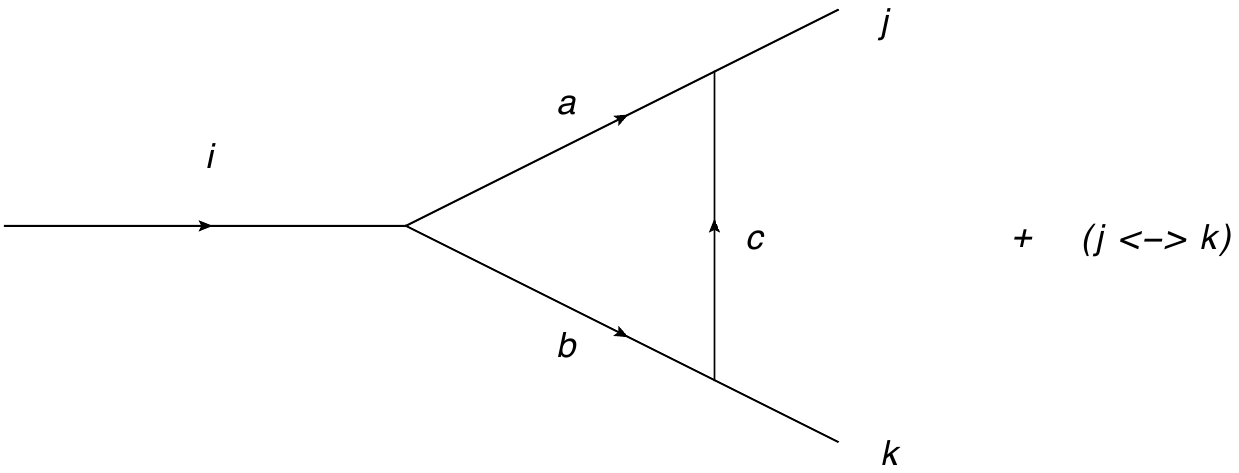,width=8cm,height=10cm}
\end{center}
\caption{One loop 3-point function} 
\label{Fig2}
\end{figure}
In leading order $\epsilon$:
\ba
\label{gamma12eps}
i(\Gamma^{(1,2)}_{div;abc})_{i,jk}=- i\tilde{\alpha}' M^{\frac{\epsilon}{2}} \frac{1}{2(4\pi)^2} \frac{1}{\epsilon}   \left(    \lambda_{i,ab}  \lambda_{j,ac} \lambda_{b,ck} + \lambda_{i,ab}  \lambda_{k,ac} \lambda_{b,cj} \right)
\ea
and
\ba
\delta \lambda_{i,jk}= \frac{1}{2(4\pi)^2} \frac{1}{\epsilon}   \left(    \lambda_{i,ab}  \lambda_{j,ac} \lambda_{b,ck} + \lambda_{i,ab}  \lambda_{k,ac} \lambda_{b,cj} \right)\,.
\ea

Now we return to (\ref{betatilde}) and consider the last two terms
\ba
&&-\frac{\epsilon}{2} \delta \lambda_{i,jk} + \frac{\epsilon}{2} \lambda_{l,mn}  \frac{\partial}{\partial \lambda_{l,mn}} 
\delta \lambda_{i,jk} \nonumber\\
&&= -\frac{1}{4} \frac{1}{(4\pi)^2}  \left(    \lambda_{i,ab}  \lambda_{j,ac} \lambda_{b,ck} + \lambda_{i,ab}  \lambda_{k,ac} \lambda_{b,cj} \right) \nonumber\\
&&\hspace{1cm}+\frac{1}{4}  \frac{1}{(4\pi)^2} \lambda_{l,mn}  \frac{\partial}{\partial \lambda_{l,mn}}   \left(    \lambda_{i,ab}  \lambda_{j,ac} \lambda_{b,ck} + \lambda_{i,ab}  \lambda_{k,ac} \lambda_{b,cj} \right)
\ea
where the expression in the last line is given by
\ba
\frac{3}{4}  \frac{1}{(4\pi)^2}  \left(    \lambda_{i,ab}  \lambda_{j,ac} \lambda_{b,ck} + \lambda_{i,ab}  \lambda_{k,ac} \lambda_{b,cj} \right)\,.
\ea
Therefore the $\beta$ function becomes:
\ba
\beta_{i,jk} &=& -\frac{\epsilon}{2} \lambda_{i,jk} + \tilde{\zeta} \lambda_{i,jk} + (\gamma_{i'i} \lambda_{i',jk}+
\gamma_{j'j} \lambda_{i,j'k}+\gamma_{k'k}\lambda_{i,jk'} )\nonumber\\
&&+\frac{1}{2}  \frac{1}{(4\pi)^2}  \left(    \lambda_{i,ab}  \lambda_{j,ac} \lambda_{b,ck} + \lambda_{i,ab}  \lambda_{k,ac} \lambda_{b,cj} \right)
\ea
or more explicitly
\ba
\beta_{i,jk} &=& -\frac{\epsilon}{2} \lambda_{i,jk} + \frac{1}{N}  \frac{1}{8(4\pi)^2}\lambda_{l,mn} \lambda_{l,mn} \lambda_{i,jk}\nonumber\\
&& -\frac{1}{8}  \frac{1}{(4\pi)^2}\left( \lambda_{i',lm} \lambda_{i,lm} \lambda_{i'jk} + \lambda_{j',lm} \lambda_{j,lm} \lambda_{i,j'k}+ \lambda_{k',lm} \lambda_{k,lm} \lambda_{ijk'} \right)\nonumber\\
&&+\frac{1}{2}  \frac{1}{(4\pi)^2}  \left(    \lambda_{i,ab}  \lambda_{j,ac} \lambda_{b,ck} + \lambda_{i,ab}  \lambda_{k,ac} \lambda_{b,cj} \right)\,,
\ea
to be considered together with the constraints given in~\eqref{conditions}.

\section{$S_{N+1}$ symmetric representation} \label{qtensors}

In order to construct  QFTs  with $N$ fields and $S_{N+1}$-invariant interactions,
let us first describe a useful representation introducing a set of $N+1$ vectors $e^\sigma$ which point in the directions
of the vertices of a $N$-simplex, i.e. a simplex in ${\mathbb R}^N$. Without specifying the explicit components with respect to some given basis, the set of vectors satisfies the following properties
\begin{eqnarray}
 && e^\sigma\cdot e^{\sigma'} =
 \sum_{i=1}^{N} e_i^\sigma  e_i^{\sigma'} = (N+1) \delta_{\sigma,\sigma'}-1 \\
 &&\sum_{\sigma=1}^{N+1} e_i^\sigma =0\,, 
 \qquad
  \sum_{\sigma=1}^{N+1} e_i^\sigma  e_j^{\sigma} = (N+1) \delta_{ij}\,.
\end{eqnarray}
Indeed these relations also determine the vectors $e^\sigma$ uniquely, up to rotations and $S_q$ transformations.

It is convenient to introduce the following tensors
\be
q^{(p)}_{i_1,\dots, i_p} = \frac{1}{N\!+\!1}Q^{(p)}_{i_1,\dots, i_p}  \quad,\quad  Q^{(p)}_{i_1,\dots, i_p} =\sum_\alpha e_{i_1}^\alpha \dots e_{i_p}^\alpha  \,,
\ee
noticing that the first two orders are trivially given by
\begin{eqnarray}
 q^{(1)}_{i_1} &=& \frac{1}{N\!+\!1} \sum_\alpha e_{i_1}^\alpha  = 0 \,, \quad
 q^{(2)}_{i_1 i_2} =\frac{1}{N\!+\!1}  \sum_\alpha e_{i_1}^\alpha e_{i_2}^\alpha  = \delta_{i_1 i_2}\,. \label{relq1and2}
\end{eqnarray}

We can compute in general contractions of product of $q^{(p)}$ tensors of various orders appearing in the diagrammatic expansion with a recursive algorithm for reduction of the tensors $q^{(p)}$~\cite{Codello:2018nbe}, without using a specific realization of the $e^{\sigma}$ vectors and $q$-tensors.

The algorithm uses the consecutive iteration of the fusion rule of two $q$-tensors
\begin{eqnarray} \label{fusion}
 q^{(p)}_{i_1,\dots,i_{p-1},j}\,  q^{(r)}_{i_{p},\dots,i_{p+r-2}}{}^{j} &=& q^{(p+r-2)}_{i_1,\dots , i_{p+r-2}} - q^{(p-1)}_{i_1,\dots,i_{p-1}}\,  q^{(r-1)}_{i_{p},\dots,i_{p+r-2}},
\end{eqnarray}
and the trace rule
\begin{eqnarray} \label{trace}
 q^{(p)}_{i_1,\dots,i_{p-2},j}{}^j &=& N\, q^{(p-2)}_{i_1,\dots,i_{p-2}}.
\end{eqnarray}
While the fusion rule generally increases the index of the fused $q$-tensor as $p \times r \to p+r-2$,
the traces of the indices of the Feynman diagrams ensure that the trace rule reduces the order as $p\to p-2$.
The procedure terminates whenever the cases $q^{(1)}_i=0$ and $q^{(2)}_{ij}=\delta_{ij}$ are encountered.

For our needs we have to compute the following two cases.
The first is given by product of two $q^{(3)}$:
\be \label{qq}
q^{(3)}_{i_1,j_1,j_2} \,q^{(3)}_{i_2}{}^{j_1,j_2} = q^{(4)}_{i_1,j_1,i_2}{}^{j_1} - q^{(2)}_{i_1,j_1}q^{(2)}_{i_2}{}^{j_1} 
= (N-1)\,\delta_{i_1,i_2}, 
\ee
where in the first equation we have fused $3\times 3 \to 4$, and in the second equation we have traced $4 \to 2$ and fused $2\times 2 \to 2$.  

The second quantity we need is the product of three copies of $q^{(3)}$:
\begin{eqnarray}
 q^{(3)}_{i_1,j_1}{}^{j_2} \,q^{(3)}_{i_2,j_2}{}^{j_3} \, q^{(3)}_{i_3,j_3}{}^{j_1}
 &=&
  q^{(3)}_{i_1,j_1}{}^{j_2} \, q^{(4)}_{i_2,j_2,i_3,}{}^{j_1} - q^{(3)}_{i_1,i_2,i_3} \nn\\
 &=&
  q^{(5)}_{i_1,j_1,i_2,i_3}{}^{j_1} - 2 q^{(3)}_{i_1,i_2,i_3}   = (N-2)\, q^{(3)}_{i_1,i_2,i_3}. \label{qqq}
\end{eqnarray}

 
\end{document}